\RequirePackage{fix-cm}
\documentclass[smallextended]{svjour3}       
\smartqed  
\usepackage[english]{babel}
\usepackage{hyperref}
\usepackage{bbold}
\usepackage{eucal}
\usepackage{soul}
\usepackage{float}
\usepackage{units}
\usepackage[T1]{fontenc}
\usepackage[utf8,applemac]{inputenc}
\DeclareUnicodeCharacter{2010}{-}
\usepackage{caption}

\usepackage[pdftex]{graphicx}
\usepackage{color}
\usepackage[toc,page]{appendix}

\usepackage{amsmath}
\usepackage{amssymb}
\usepackage{mathtools}
\usepackage{braket}

\usepackage[dvipsnames]{xcolor}

\newcommand{\SOT}{\operatorname{SO}(3)}

\begin{document}

\title{$\mathcal{PT}$-symmetry in compact phase space for a linear 
  Hamiltonian
}


\author{Ivan F. Valtierra         \and Mario B. Gaeta \and Adrian Ortega \and Thomas Gorin 
}


\institute{Departamento de F\'isica, Universidad de Guadalajara, Blvd. Gral. Marcelino Garc\'ia Barrag\'an 1421, C.P. 44430, Guadalajara, Jalisco, M\'exico.
}

\date{Received: date / Accepted: date}

\maketitle

\begin{abstract}
\noindent We study the time evolution of a $\mathcal{PT}$-symmetric, non-Hermitian
quantum system for which the associated phase space is compact. We focus on the
simplest non-trivial example of such a Hamiltonian, which is linear in the 
angular momentum operators. In order to describe the evolution of the system,
we use a particular disentangling decomposition of the evolution operator, 
which remains numerically accurate even in the vicinity of the Exceptional 
Point. We then analyze how the non-Hermitian part of the Hamiltonian affects 
the time evolution of two archetypical quantum states, coherent and Dicke 
states. For that purpose we calculate the Husimi distribution or $Q$ function and
study its evolution in phase space. For coherent states, the characteristics
of the evolution equation of the Husimi function agree with the trajectories of the
corresponding angular momentum expectation values. This allows to consider 
these curves as the trajectories of a classical system. For other types of
quantum states, e.g. Dicke states, the equivalence of characteristics and
trajectories of expectation values is lost.
\end{abstract}

\section{Introduction and motivation}
\label{sec:Introduction}

\noindent Non-Hermitian quantum systems may still have a real 
spectrum, if they are $\mathcal{PT}$-symmetric~\cite{Benderbook2018}. Since
their introduction by Bender and Boettcher in 1998~\cite{BenderPRL1998}, 
these systems have found a wide range of applications~\cite{HeissJPA2012,BenderJP2015,FengNature2017,El-GanainyNature2018,Miri2019}. One of the defining 
features of such systems is that the corresponding Hamiltonian may have real or
complex eigenvalues. In the first case, one says that the 
$\mathcal{PT}$-symmetric phase is intact, in the second that it is broken.
The transition between these phases occurs at so-called ``Exceptional 
Points'' (EPs), at which two or more eigenvalues and eigenvectors coalesce. In 
this case, the Hamiltonian becomes defective~\cite{Katobook1995}. EPs have been
related to many remarkable phenomena. Some recent examples 
are chirality~\cite{PhysRevLett.86.787,Mailybaev2005,Peng2016}, unidirectional 
invisibility~\cite{Regensburger2012,Lin2011,Feng2012}, enhanced 
sensing~\cite{Chen2017} and the possibility to stop 
light~\cite{GoldzakPRL2018}.  
From the transport point of view, a $\mathcal{PT}$-symmetric Hamiltonian
offers a convenient way to describe a physical system with gains and 
losses.
Such phenomena can be studied conveniently in tight-binding Hamiltonians~\cite{JinPRA2009,YogeshPRA2010,OrtegaJPA2020}.

Only recently, $\mathcal{PT}$-symmetric  systems have been studied from a 
semiclassical perspective, 
also~\cite{GraefeJPA2010,GraefePRA2011,PraxmeyerPRA2016}. In this context it is
natural to use phase space representations of the quantum system in question.
These investigations have been focused mainly on systems described by the 
Heisenberg-Weyl algebra where the associated phase space is a two-dimensional
plane. The case of a compact phase space, which corresponds to the angular
momentum algebra has received much less attention, though 
note~\cite{GraefeJPA2010,GrKoNi10,MudGra20}. 

In this work, we provide a thorough analysis of the simplest non-trivial
$\mathcal{PT}$-symmetric system, where the Hamiltonian is a linear function of
the angular momentum operators. For such systems, the Bloch sphere plays the
role of a classical phase space.
In the case of a linear Hermitian Hamiltonian, the phase space representations 
(for instance Glauber, Wigner or Husimi) \cite{refI1,refI2,refI3,refI4} of the quantum system evolve along characteristics, conserving their function 
values -- just as in the case of the classical Liouville equation. As a 
consequence, the expectation values of the angular momentum operators follow 
these characteristics also, and this allows one to identify them as classical 
trajectories. 


In the case of a non-Hermitian and $\mathcal{PT}$-symmetric Hamiltonian, the 
situation is different: The evolution equations for the different quantum
representations in phase space do no longer agree, and some of them may no 
longer be solved in terms of characteristics.

In this work we use the Husimi function, 
which can still be solved with the method of characteristics. There the only 
difference is that the function changes its value along the characteristics, 
which is equivalent to the existence of sources and sinks in the phase space. 
The expectation values of the angular momentum operators follow different 
trajectories, lying inside the Bloch ball, which depend on the complete shape 
of the quantum states. Surprisingly, it is still possible to return to the old
unified picture. To achieve this, one has to limit to the evolution
of coherent states. These states conserve their shape and follow the
characteristcs of the evolution equation for the Husimi function. However, they
do not conserve their norm. 
This allows one to speak of a corresponding classical system, where the 
trajectories represent localized excitations which may vary in 
intensity~\cite{MudGra20}, according to the sources and sinks present in the
system. 

\noindent The paper is organized as follows. In Sec.~\ref{sec:Model}, we describe our 
model and introduce the quantities of interest: the evolution operator,
expectation values and the Husimi function (or $Q$ function). In 
Sec.~\ref{sec:tevolH} we present two methods to compute the evolution of the 
system, the diagonalization of the Hamiltonian with a non-unitary similarity 
transformation and the disentangling decomposition inspired by the 
decomposition of a general rotation using Euler angles. In Sec.~\ref{EvoPhaSpa} 
we discuss our results for the time evolution of the quantum system in phase
space. After this we consider the Husimi function and the evolution of expectation values
in Sec. \ref{sec:Conclusions}. In the appendix, we collect a few general 
properties of expectation values and variances for coherent states and Dicke
states (in App.~\ref{sec:App1}), derive the evolution equation for the Husimi function (in App.~\ref{EvoPS}) and the analytical solution to the 
Ehrenfest equation for the angular momentum expectation values (in 
App.~\ref{AppSOL}). 

\section{General definitions and the model}
\label{sec:Model}

\noindent We are interested in systems where the dynamical symmetry group is 
$\SOT$.  
The angular momentum operators $\lbrace{S_x,S_y,S_z\rbrace}$ are the 
generators of the corresponding Lie algebra $\mathfrak{so}(3)$, and hence 
fulfill the commutation relations $[S_x,S_y] = i S_z$ (and cyclic permutations 
of it). For simplicity, we set $\hbar=1$. 
The eigenvectors of the operator $S_z$ are the Dicke states 
$\left|S, m \right\rangle$, where
\begin{equation}
S_z \left| S,m \right\rangle = m \left| S,m \right\rangle \quad ,\qquad 
   -S\leq m \leq S
\label{Szbase}\end{equation}
with $S$ being a positive integer. 
As the generator of the dynamics, we choose the simplest non-trivial, 
linear, PT-symmetric Hamiltonian,
\begin{equation}
  H = 2v\; S_x - 2i\, \gamma\; S_z = H_0 + i\; \Gamma\; ,
  \label{eq:H}\end{equation}
with real parameters $\gamma$ and $v$, such that both operators
$H_0$ and $\Gamma=-2\gamma S_z$ are Hermitian. 
This Hamiltonian can describe, under a Schwinger transformation, 
the motion of a non-interacting Bose-Einstein condensate in a two-well 
potential under $\mathcal{PT}$-symmetry~\cite{GraefePRL2008}. 

The Hamiltonian in Eq.~(\ref{eq:H}) is $\mathcal{PT}$-symmetric. 
This means that there exist an antilinear (conjugated linear) operator $\mathcal{T}$ 
and a linear involution $\mathcal{P}$ (i.e. $\mathcal{P} = \mathcal{P}^{-1}$) 
which commute, $[\mathcal{T},\mathcal{P}] = 0$, such that 
$[\mathcal{PT},H] = 0$~\cite{MudGra20,JonMat10}. 
As a consequence the Hamiltonian may have real eigenvalues.

In our case, the parity operator may be defined by its action on the Dicke 
states:
\begin{equation}
\forall\; -S \le m\le S\quad :\quad \mathcal{P}\; |S,m\rangle = |S,-m\rangle 
\; ,
\end{equation}
and the time-reversal operator by its action on an arbitrary state written as a
linear combination of the Dicke states:
\begin{equation}
\mathcal{T}\; \sum_m c_m\; |S,m\rangle = \sum_m c_m^*\; |S,m\rangle \; .
\end{equation} 
The operators $\mathcal{P}$ and $\mathcal{T}$ fulfill all the requirements 
mentioned above. In addition it holds that $\mathcal{T}^2 = +\, \mathbb{1}$ 
such that we may speak of an ``even'' $\mathcal{PT}$ symmetry. 

\subsection{Evolution and expectation values}

\noindent We find the solution of the Schr\" odinger equation
\begin{equation}
i\, \frac{d}{dt}\; |\Psi\rangle = H\; |\Psi\rangle \, 
\end{equation}
in terms of the non-unitary evolution operator
\begin{equation}
U(t) = e^{-i H t}\; .
\label{DefEvolU}\end{equation}
Accordingly, we obtain for the evolution of the corrresponding density matrix
\begin{equation}
\varrho(t) = U(t)\; \varrho(0)\; U(t)^\dagger \; .
\end{equation}
This shows that the density matrix $\varrho(t)$ does not remain normalized in
general. However, $\rho\left(t\right)$ is Hermitian by construction, so its trace, and also the expectation values of observables calculated from it, are real.

Non-Hermitian Hamiltonians are typically used to describe quantum systems or 
wave-mechanical systems with gains and losses. In the single particle case,
this implies that ${\rm tr}[ \varrho(t)\, ] \le 1$ is the probability that
the measurement of some observable $\hat A$ actually finds the particle and 
gives a valid result. In the case of many particles (e.g. a Bose-Einstein 
condensate) or a classical wave field, the density matrix describes the 
intensity of the field as it varies over time. In both cases, we calculate
the expectation value of $\hat A$ using the density matrix as
\begin{equation}
\langle \hat A\rangle 
 = \frac{{\rm tr}[\varrho(t)\, \hat A\, ]}{{\rm tr}[\varrho(t)\, ]} \; .
\label{ExpVal}
\end{equation}
This is because in the single particle case, measuring $\hat A$ implies to 
average over many realizations of the measurement. Then, the only meaningful 
average is that over those realizations where the measurement was actually 
successful.

\subsection{Husimi function}


\noindent A useful method to visualize a quantum system consists in mapping its 
state $\varrho$ into a complex valued function defined on the corresponding 
classical phase space. According to Weyl~\cite{Weyl} and later
Stratonovich and Moyal~\cite{Stratonovich,moyal} there
are several choices for such a mapping, which can be tagged by the 
parameter $s\in \{ +1,0,-1\}$. In that case, the different values correspond
to the $P$ (Glauber), $W$ (Wigner) and $Q$ (Husimi) functions, respectively. 
Since the dynamical symmetry group is $\SOT$, the classical phase space is the 
so-called Bloch sphere $\mathbb{S}^{2}$. 
For our purposes, we will use the
(real positive) Husimi 
function, which is defined as
\begin{equation}
Q_\varrho(\theta,\phi) =\langle \theta ,\phi |\varrho|\theta, \phi \rangle 
  = {\rm tr}\big [\, \varrho\, \hat{\omega}_Q(\theta,\phi)\, \big ]\; ,
\end{equation}
where the angles $(\theta,\phi)$ parametrize the points in $\mathbb{S}^{2}$ and 
\begin{equation}
|\theta,\phi\rangle = e^{-i \phi S_z} e^{-i \theta S_y} \left| S,S \right\rangle
\label{edocoh}
\end{equation}
is the coherent state \cite{refcoh} with angular momentum expectation
values
\begin{equation}
\begin{pmatrix} \langle S_x\rangle\\ \langle S_y\rangle\\ \langle S_z\rangle
\end{pmatrix} = S\begin{pmatrix} 
  \sin\theta\, \cos\phi\\ \sin\theta\, \sin\phi\\ \cos\theta \end{pmatrix} \; .
\end{equation}
The kernel operator
$\hat{\omega}_Q(\theta,\phi) = |\theta,\phi\rangle \langle \theta,\phi|$ can be
written as
\begin{equation}\label{kernel}
\hat{\omega}_Q(\theta,\phi) = \frac{2\sqrt{\pi}}{\sqrt{2S+1}} \sum_{L=0}^{2S}\sum_{M=-L}^{L}Y^{*}_{L,M}(\theta,\phi) \hat{T}^{(S)}_{L,M}C^{SS}_{SS;L0}\; ,
\end{equation}
where $Y_{L,M}(\Omega)=(-1)^{M}Y^{*}_{L,-M}(\Omega)$ is the standard spherical
harmonic function. In Eq.~\eqref{kernel} we used the irreducible tensor 
operators 
\begin{equation}
\hat{T}^{(S)}_{L,M} = \sqrt{\frac{2L+1}{2S+1}}\sum_{m,m^{'}=-S}^{S}C^{S,m^{'}}_{S,m;L,M}|S,m^{'}\rangle\langle S,m|.
 \end{equation}
Here $C^{S,m^{'}}_{S,m;L,M}$ and $C^{SS}_{SS;L0}$ are the Clebsch-Gordan 
coefficients~\cite{Varshalovichbook}. 
The Husimi function is real and positive.

\begin{figure}
 \centering
 \hspace*{-2mm}
\includegraphics[width=0.68\textwidth]{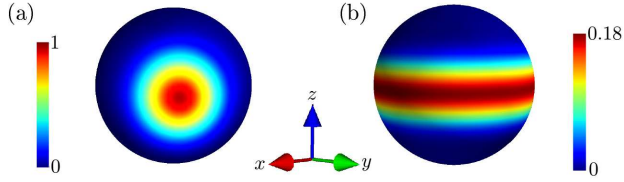}
\caption{(a) Husimi function $Q_{|\theta_0,\phi_0\rangle}$ of the coherent state with $\theta_0=\pi/2, \phi_0= \nicefrac{\pi}{4}$. (b) Husimi function of the Dicke state $|S=10,m=0\rangle$.}
\label{estcoh}\end{figure}

\noindent In order to study the dynamics generated by $H$, 
we choose as initial conditions two archetipycal states: a coherent state 
(semiclassical) and a Dicke state (no semiclassical). The Husimi function of a
coherent state $|\theta_0,\phi_0\rangle$ is
\begin{eqnarray}
Q_{|\theta_0,\phi_0\rangle}= \Big(  \frac{1+ \mathbf{n_0}\cdot \mathbf{n}}{2} 
  \Big)^{2S},
\end{eqnarray}
which is a localized distribution in phase space, centered at the point 
$\mathbf{n_0}=(\sin\theta_0 \cos\phi_0,\sin\theta_0 \sin\phi_0,\cos\theta_0)$ 
with the variables $(\theta,\phi)$ contained in 
$\mathbf{n}=(\sin\theta \cos\phi,\sin\theta \sin\phi,\cos\theta)$. 
Fig.~\ref{estcoh}(a) shows an example for $\theta_0 =\pi/2, \phi_0 = \pi/4$.

On the other hand, the Husimi function of a Dicke state is usually not 
localized,
\begin{equation}
Q_{|S,m \rangle} = \frac{1}{\sqrt{4\pi}} \sum_{L=0}^{2S}
   \sqrt{2L+1}\; 
Y^{*}_{L0}(\theta,\phi)\; C^{SS}_{SS;L0}C^{S m}_{S m; L0}\; .
\end{equation}
For instance, the Husimi function of the state $\left| S,0 \right\rangle$, 
is delocalized in the $\phi$-coordinate, as can be seen in 
Fig~\ref{estcoh}(b). General expressions for expectation values and variances 
of coherent states and Dicke states are collected in App.~\ref{sec:App1}.

\section{Decomposition of the evolution operator}
\label{sec:tevolH}

\noindent 
In this section, we consider two methods to calculate the evolution operator
$U(t)$ from Eq.~(\ref{DefEvolU}) analytically. The first is based on a 
similarity transformation which diagonalizes $H$. The second, so-called 
``disentangling method'', factorizes the evolution into three elementary parts.

\subsection{Diagonalization method} \label{secDiag}

\begin{figure}
\centering
  \hspace*{-3mm}
\includegraphics[width=0.7\textwidth]{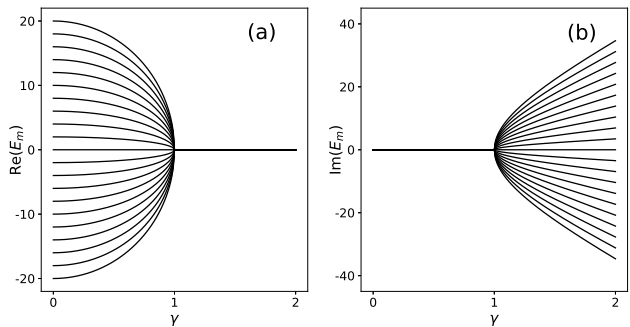}
\caption{The eigenvalues $E_m$ of the Hamiltonian $H$, as given in 
Eq.~(\ref{Heigenvals}) as a function of $\gamma$ for $S=10$ and $v=1$. In panel
(a) the eigenvalues are real for $\gamma < v$, in panel (b) the eigenvalues are 
imaginary for $\gamma > v$. As $\gamma\to v$, all eigenvalues coalesce in a 
single EP.}
\label{fig:1}\end{figure}

\noindent Following~\cite{GraefeJPA2008}, the Hamiltonian can be rewritten as
\begin{equation}
  H = e^{\alpha S_y}\; \tilde H\; e^{-\alpha S_y}\; , \qquad
  \tilde H= -2\sqrt{v^2 - \gamma^2}\; S_z\; ,
  \label{eq:H2}
\end{equation}
with $\cosh\alpha = \gamma / \sqrt{\gamma^2 - v^2}$ and 
$\sinh\alpha=v/\sqrt{\gamma^2-v^2}$. 
Eq.~(\ref{eq:H2}) shows that the eigenvalues of $H$ are
given by 
\begin{equation}
E_m = - 2m\sqrt{v^2 - \gamma^2}\; , \qquad -S \le m\le S \; .
\label{Heigenvals}\end{equation}
The behavior of the eigenvalues is shown in Fig.~\ref{fig:1}. Clearly, they are 
all real as long as $\gamma < v$, they coalesce for $\gamma = v$ in one single 
point, and become all imaginary when $\gamma > v$. Thus our system has a unique
exceptional point at $\gamma = v$.

\begin{figure*} 
\centering
\hspace*{-0.3mm}
\includegraphics[width=0.75\textwidth]{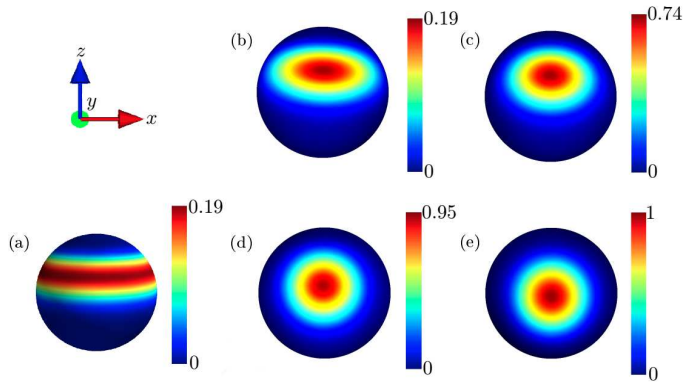}
\caption{Initial Dicke state under the transformation $e^{-r S_y}$. The initial 
Dicke state $|\psi\rangle=|S=10,m=4\rangle$ (a)  is deformed as a 
function of $\gamma$: In (b) we show the 
deformation for $\gamma = 0$.2, in (c) for $\gamma=0.5$, in (d) for 
$\gamma=0.9$ and in (e) for $\gamma=0.99$. Interestingly, for 
$\gamma \approx 1$ the transformation turns a Dicke state into a coherent one.}
\label{TrSimD}
\end{figure*}

The dynamics of the system is very different for $\gamma < v$ and
$\gamma > v$. 
On the one hand, when $\gamma > v$, we obtain 
$\tilde H = -2i\, \sqrt{\gamma^2 - v^2}\; S_z$ such that
\begin{equation}
U(t) = e^{\alpha\, S_y}\; e^{-2\sqrt{\gamma^2 - v^2}\, S_z\, t}\;
   e^{-\alpha\, S_y}\; ,
\end{equation}
with $\alpha = {\rm atanh}(v/\gamma)$, real. Thus the evolution operator 
consists entirely of exponential solutions -- i.e. rotations with purely
imaginary arguments. Therefore, for sufficiently long times, $U(t)$ maps any
initial state $|\Psi_0\rangle$ to
\begin{equation}
\lim_{t\to\infty} U(t)\; |\Psi_0\rangle \propto e^{\alpha\, S_y}\; |S,-S\rangle
\; .
\end{equation}
On the other hand, for $\gamma < v$, the eigenvalues of $H$ are all real, 
and the evolution in time consists of rotations around the $z$-axis. For the 
similarity transformation, we find $\alpha = -i\pi/2 + r$, where
\begin{equation}
 r = {\rm atanh}(\gamma/v) = 
   \frac{1}{2}\, \ln\Big (\, \frac{v + \gamma}{v - \gamma}\, \Big )\; .
\end{equation}
In this way, we arrive at
\begin{align}
U(t) &= e^{-i \pi/2\, S_y}\, e^{r\, S_y}\; 
  e^{-2i\sqrt{v^2 - \gamma^2}\, S_z\, t}\; e^{-r\, S_y}\, e^{i\, \pi/2\, S_y}
\notag\\
 &= e^{r\, S_y}\; e^{-2i\sqrt{v^2 - \gamma^2}\, S_x\, t}\; e^{-r\, S_y}\; .
\label{eq:Ut}\end{align}
According to this equation, the evolution of a quantum state consists of a 
rotation around the $x$-axis, sandwitched between a deformation of the state 
with the non-unitary operator $e^{-r\, S_y}$, and its inverse. In what follows 
we limit ourselves to this regime ($\gamma < v$), where some unitary dynamics 
(the rotation) still persists.

In Fig.~\ref{TrSimD}, we show the effect of the deformation operator 
$e^{-r\, S_y}$ when choosing a Dicke state as initial state. For that purpose,
we plot the Husimi function of the state before [panel (a)] and after applying 
the deformation operator.  The remaining results are shown in the rest of
the panels for increasing values of $\gamma$ (see the figure for details). This deformation changes the norm of the 
state very strongly, in particular when $\gamma$ comes close to $v$. For that
reason we plot the normalized Husimi functions.

Note that for $r$ sufficiently large (i.e. $\gamma$ sufficiently close to $v$),
the operator $e^{-r S_y}$ will map practically any state onto its eigenstate
corresponding to the largest negative eigenvalue. This is a coherent state
placed on the $y-$axis. 
This explains the fact that for increasing value of $\gamma$ [panels (b),
  (c), (d), (e)] the deformed state becomes ever more similar to the just
mentioned coherent state.

\subsection{Decomposition method}

\noindent In the interval $\gamma \in [0,1)$, the spectrum of $H$ is real but the 
associated eigenfunctions tend to align to each other as 
$\gamma \rightarrow 1$, i.e. close to the EP. Numerically, this yields 
inacuracies which we want to avoid. 
For that purpose, we follow \cite{weigert}, which allows to 
disentangle the evolution operator as follows.
In general, the disentangling method searches for typically time-dependent
coeficients 
$\alpha, \beta, \gamma$ such that
\begin{equation}
e^{-i H t} = e^{-i\alpha\, S_1}\; e^{-i\beta\, S_2}\; 
   e^{-i\gamma\, S_3}\; .
\label{Habc}
\end{equation}
where the operators $S_1, S_2, S_3$ 
are chosen from the angular momentum operators $S_x, S_y, S_z$. Many different
combinations are possible~\cite{Varshalovichbook}, but as long as the dynamics 
is unitary, one refers to the parameters $\alpha, \beta, \gamma$ as Euler 
angles. In the present case, however, the dynamics is not unitary and some of 
the parameters must be complex or imaginary. After considering and discarding
the quantum mechanical $S_z$--$S_y$--$S_z$ scheme, as well as the Gauss
decomposition scheme~\cite{Varshalovichbook} due to eventual 
singularities, we found the following decomposition particularly convenient:
\begin{equation}
  U(t) = e^{-i \alpha S_z}e^{-i \beta\, S_y}e^{-i h\, S_x} \; .
\label{disec}\end{equation}
Deriving both sides of (\ref{disec}) by $t$, and then comparing them 
(factoring out the evolution operator), using the lineal independence of $S_x$, 
$S_y$ and $S_z$, one arrives at the following system of differential 
equations:
\begin{align}
\dot\alpha - \dot h\; \sin\beta &= -2i\, \gamma\notag\\
\dot\beta\; \cos\alpha + \dot h\; \sin\alpha\; \cos\beta &= 0\notag\\
\dot\beta\; \sin\alpha - \dot h\; \cos\alpha\; \cos\beta &= -2v \; .
\end{align}
This system of equations can be converted into a real system for real parameters
$f= i\alpha, g= i\beta, h$ which yields
\begin{align}
\dot f &= \tanh(g)\; \cosh(f)\; 2v + 2\gamma\notag\\
\dot g &= -2v\; \sinh(f)\notag\\
\dot h &= 2v\; \frac{\cosh(f)}{\cosh(g)}\; .
\label{eq:syseq}\end{align}
The initial condition is $U(0) = \mathbb{1}$, which implies 
$f(0) = g(0) = h(0) = 0$. Therefore, the system of equations has a 
unique solution, which assures that $f,g,h$ are real functions in time. In
principle the system of ordinary differential equations can be solved
analytically, by calculating the eigenvalues and eigenstates of the coefficient
matrix. For simplicity, however, we have used numerical solutions of
Eq.~\ref{eq:syseq} and the results are presented below.


\begin{figure}
  \centering
 \hspace*{-3mm}
  \includegraphics[width=0.6\textwidth]{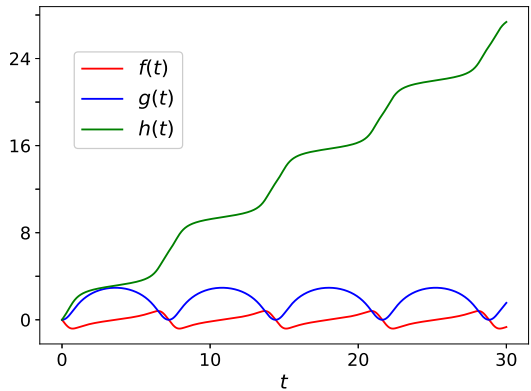}
  \caption{Numerical solutions of Eq. (\ref{eq:syseq}) for
    $v=1$ and $\gamma = 0.9$.}
\label{fig:2}
\end{figure}

\noindent In Fig. \ref{fig:2} we show the numerical solutions for these functions, for 
$v=1$ and $\gamma = 0.9$, in a time interval $t\in[0,30]$. 
The function $h(t)$ (green) represents the way the unitary
operator $\exp(-i h(t)S_x)$ acts on the initial state. This only generates 
rotations around the $x$-axis on the phase space. Further, $h(t)$ is a 
monotonously increasing function. On the other hand, $f(t)$ (red) and $g(t)$ (blue) are 
periodic functions, such that the corresponding evolution operators perform
imaginary rotations, which have their origin in the non-unitary part of the Hamiltonian
$H$. 
This will lead to interesting dynamical effects discussed below.
The dependence on $\gamma$ is reflected on the slope of $h(t)$ and the periods 
of $f(t)$ and $g(t)$. 

In fact, according to Eqs. (\ref{eq:H2}) and (\ref{eq:Ut}) 
one can see that the evolution is periodic with period 
$T=\pi/\sqrt{1-\gamma^2}$ ($v=1$). This is indeed the period of $f(t)$ and 
$g(t)$. For the monotone function $h(t)$ and the corresponding operator 
$\exp(-i h(t)S_x)$, this means that $h(nT) = 2 \pi n$, for integer $n$. So, 
applying a linear fit to $h$  yields the slope $2\sqrt{1-\gamma^2}$.

\subsection{Time evolution of the quantum state intensity}
\label{TEHS}

In our system the norm of a evolving quantum state is not conserved. From 
the physical point of view such a state may represent a large number of 
particles in a (may be effective) single particle quantum state (e.g. 
Bose-Einstein condensate~~\cite{GraefePRL2008}, or quasi-particles in quantum transport~\cite{Feng2012}) or the 
quantum state can be interpreted as a classical wave field~\cite{PhysRevLett.86.787,RochaPRApp2020}. In any case, the change 
in the norm squared of the state, i.e. its trace, can be interpreted as the loss or gain of 
particles or wave field intensity. The term ``quantum state intensity'' should 
be understood in this manner.

\begin{figure}
 \centering
 \hspace*{-8mm}
 \includegraphics[width= 1.1\textwidth]{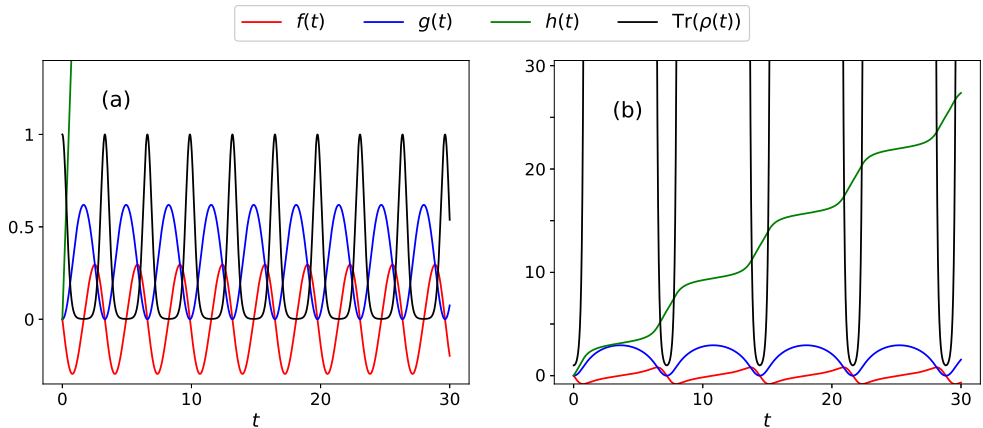} 
\caption{Time evolution of the functions involved in the disentagling
of the evolution operator Eq. (\ref{disec}) along with the time evolution
of the trace for the initial state 
$\left|\theta_0 = \nicefrac{\pi}{2}, \phi_0 = \nicefrac{\pi}{4} \right\rangle$;
$\sin \phi_0 \approx 0$.$707$. (a) $\gamma = 0.3$, (b) $\gamma = 0.9$.}
\label{reg1tr}\end{figure}

At this point it is instructive to analyze the evolution of the trace of the
two different initial states given in the introduction along with the
decomposition of the evolution operator, Eq.~(\ref{disec}). The intensity of
each state oscillates in different ways, which depend on the initial state
and also on the value of $\gamma$. 
If the initial coherent state is centered in the $(S_x,S_y)$ plane, i.e. $\theta_0=\pi/2$, we observe that, for $\gamma < \sin \phi_0$, the trace oscillates in the values
$0<\rho(t)\leq 1$. The minimal value that the trace attains depends on the
initial state and $\gamma$. Further, the minima of the trace corresponds to
the maxima of $g(t)$. These effects are shown in Fig.~\ref{reg1tr}(a) for $\sin \phi_0 \approx 0$.$707$, and $\gamma = 0.3$.

\noindent For $\gamma > \sin \phi_0$, the trace
oscillates between $1\leq\operatorname{Tr}(\rho(t))<M$, for $M>1$, where the
maximum value of $M$ can be quite large. (This factor of $\sin \phi_0$ emerges when analyzing the evolution 
in phase space: Sec.~\ref{TDQD}.) In Fig.~\ref{reg1tr}(b) for
$\gamma = 0.9$, clearly close to the EP, this value of $M$ can be as large
as $\sim 10^9$. Besides, the minima of the trace corresponds to the minima of $g(t)$. This 
behaviour is also exhibited for an initial Dicke state with $m=0$, but in contrast 
to the previous case, it holds for all values of $\gamma$.

\begin{figure}
  \centering
  \hspace*{-3mm}
  \includegraphics[scale=0.55]{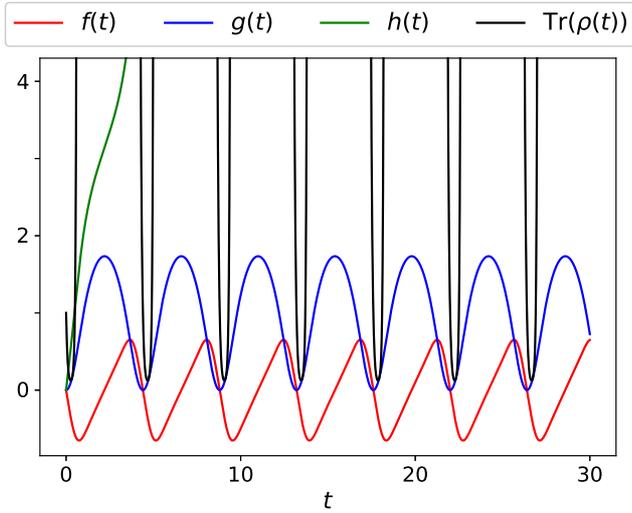}
  \caption{Time evolution of the functions involved in the disentagling
    of the evolution operator Eq. (\ref{disec}) along with the time evolution
    of the trace for the initial state $|S=10,m=4\rangle$ and $\gamma=0.7$.}
  \label{cond3}
\end{figure}

\noindent If the initial coherent state is not centered in the $xy$ plane, or for initial Dicke states with $m\neq 0$,
the trace oscillates between a minimum value (that for some cases can be
zero) and a value greater than one, irrespectively of the value of $\gamma$.
As in the previous cases, this
maximum value can be quite large. This generic behaviour is shown
in Fig. \ref{cond3} for a Dicke state $\left| S = 10, m= 4 \right\rangle$
and $\gamma=0.7$. Further, notice that the minima of the trace do not
coincide, though is close, to
 the minima of $g(t)$ (c.f. Fig.~\ref{cond3}).

\section{Time evolution in phase space} \label{EvoPhaSpa}

\noindent In this section, we consider the evolution of the Husimi distribution and we analyze the evolution of the angular momentum expectation
values. In the case of initial coherent states, we find a one-to-one 
correspondence between the trajectories traced by the expectation values and
the characteristics of the evolution equation for the Husimi function. This is due
to the fact that coherent states remain coherent during evolution -- only their
intensity (trace) changes. For initial Dicke states, this is no longer true.

\subsection{Time dependent Husimi distribution } \label{TDQD}

\noindent We are now in the position to study the time evolution in phase space of the system. 
Using the decomposition from Eq.~(\ref{eq:H}) of the Hamiltonian into a 
Hermitian and an anti-Hermitian part, $H = H_0 + i\Gamma$, the von Neumann 
equation can be written as 
\begin{equation}
  i \partial_t \rho(t) = [H_0,\rho(t)] + i[\Gamma, \rho(t)]_{+}.
  \label{eq:rho}
\end{equation}
Following App.~\ref{EvoPS}, one gets the equation of motion of the 
Husimi function 
\begin{equation}
i \partial_t{Q}(\theta,\phi) =  -2(v\hat{l}_x +2 i\gamma S\cos\theta -i\gamma \sin\theta \partial_\theta )Q(\theta,\phi),   \label{eq:W} 
\end{equation}


\noindent where 
$\hat{l}_x = i (\sin\phi \partial_\theta+\cot\theta \cos\phi \partial_\phi)$. The method of characteristics then yields
\begin{eqnarray}\label{cleq}
\dot{\theta} &= &-2 v \sin\phi + 2\gamma \sin\theta, \\ \notag
\dot{\phi} &= &-2v \cot\theta \cos\phi,\\ \notag
\dot{Q} &=& 4\gamma S \cos\theta.
\end{eqnarray}

\noindent It is worth noting that the first two  equations above are consistent with the 
Ehrenfest equations derived for coherent states (Sec.~\ref{SecExV}, Eq.~(\ref{SEE})) and the last equation 
describes the deformation of the distribution along the characteristics.

As we have demonstrated in the App.~\ref{AppSOL}, the trajectories of the expectation values (in the case of coherent states) or the characteristics of Eq.~(\ref{cleq}), trace circles in phase space symmetric respect to deflection on the $yz$ plane. This implies that for an initial state with a Husimi function of the same symmetry that this symmetry is conserved. So for instance, as $\left\langle S_x \right\rangle = 0$ for a Dicke state, $\left\langle S_x \right\rangle$ will remain zero for all times (See Fig.~\ref{figtray}(b)).   

\noindent We continue with the analysis of the time evolution of the Husimi function
for $\gamma \in[0,1)$ in phase space. We observe
that the distribution only spins
around for coherent states (though there is an exception which is
explained below in this section) but for initial Dicke states beside spinning it deforms too (also see below in this section). As before, the period and amplitude of the rotation
depend on the initial state, as well as on the value of $\gamma$. The rotation
period is the same as the periods of $f(t)$ and $g(t)$.

\begin{figure}[htpb]
  \centering
  \hspace*{-1.5mm}
  \includegraphics[scale=0.3]{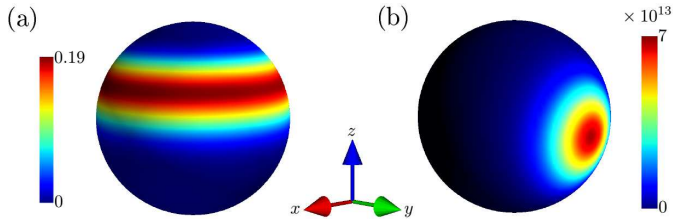}
  \caption{Time evolution of an initial Dicke state in phase space for $\gamma=0.7$. 
   (a) the state $\left|S=10,m=4 \right\rangle$. $\gamma = 0.7$ at $t=0$. (b) the evolved state at $t=2.4$.}
\label{Dicke2}
\end{figure}

\noindent An interesting behaviour occurs on the rotating distribution. When it comes closer to the point $\left(\theta = \nicefrac{\pi}{2} , \phi = \nicefrac{\pi}{2} \right)$, the distribution moves slower than it does on other regions. This also happens with initial Dicke states, but besides the rotation of the
distribution, its Husimi function also becomes localized. Fig. \ref{Dicke2} shows this effect for an initial Dicke state
$|S=10,m=4\rangle$ with $\gamma=0.7$ at $t=2.4$.

\noindent On the other hand, when the Husimi distribution comes closer to the point $(\theta = \nicefrac{\pi}{2}, \phi = -\nicefrac{\pi}{2})$, while it moves faster for both types of states, also the correspondig distribution of the initial Dicke state recovers its original form.

Another courious effect happens for an initial coherent state located on the
plane $(x,y)$. As
$\gamma$ gets closer to $\sin \phi_0$, the amplitude of the rotations decreases until it becomes zero for
$\gamma = \sin \phi_0$. For this particular value of $\gamma$ the Husimi function
becomes stationary. When $\gamma >\sin \phi_0$, the amplitude increases and
we recover the spinning pattern. Before and after that value, the distribution moves initially towards different directions, so the center of the circle followed by the distribution can be on one side or the other of the starting point. An explanation for the $\sin \phi_0$ factor is given in App. \ref{AppSOL}. 


Most of this discussion is then analyzed from another perspective in Sec.~\ref{SecExV}, where it is also illustrated.

\subsection{Ehrenfest equations and the time evolution of the expectation    values} \label{SecExV}

\noindent In order to get an insight of what will be the time evolution of the
distributions in phase space, we study the first and second moments of the
operators $\lbrace S_x, S_y, S_z\rbrace$. Denoting for an observable $\hat{A}$, $\mbox{tr}(\, \hat{A} \rho\, ) = \langle\langle \hat{A}\rangle\rangle $, we rewrite Eq.~(\ref{ExpVal}) as
\begin{equation}
\langle \hat{A} \rangle 
   = \frac{\langle\langle\, \hat{A}\, \rangle\rangle}{{\rm tr}(\rho)}\; .
\label{tranorm}\end{equation}

\noindent The evolution of the expectation value of $\hat{A}$ can be obtained 
from the Ehrenfest equation. However, we have to take into account that the trace of the density matrix depends on time. For a general Hamiltonian with
$H = H_0 + i \Gamma$ with Hermitian operators $H_0$ and $\Gamma$ one 
obtains~\cite{GraefeJPA2010}
\begin{equation}
i\frac{d}{dt}\langle\!\langle A\rangle\!\rangle = \langle\!\langle [ A, H_0 ] + i [ A, \Gamma ]_+ \rangle\!\rangle.
\end{equation}
Writing for the trace of the density matrix 
${\rm tr}(\rho) = \langle\langle\mathbb{1}\rangle\rangle$, we obtain
\begin{equation}
  \frac{d}{dt}\langle\!\langle \mathbb{1} \rangle\!\rangle  
= 2\, \langle\!\langle \Gamma \rangle\!\rangle
\label{dtraza}
\end{equation}
and
\begin{eqnarray}
\frac{d}{dt}\langle A\rangle &=& 
   \frac{1}{\langle\!\langle \mathbb{1} \rangle\!\rangle}
   \frac{d}{dt}\langle\!\langle A\rangle\!\rangle 
 - \frac{\langle\!\langle A\rangle\!\rangle }
      {\langle\!\langle \mathbb{1}\rangle\!\rangle^{2}}\;
   \frac{d}{dt}\langle\!\langle \mathbb{1}\rangle\!\rangle\notag\\
&=& \frac{1}{\langle\!\langle \mathbb{1} \rangle\!\rangle}
   \frac{d}{dt}\langle\!\langle A\rangle\!\rangle
   -2\, \langle A\rangle\, \langle\Gamma\rangle \; .
 \end{eqnarray}
\noindent Applying these equations to the Hamiltonian from Eq.~(\ref{eq:H}), we find for the normalized expectation values
\begin{equation}
\begin{split}
\frac{d}{dt}\langle S_x \rangle &= -2\gamma \langle [S_z,S_x]_{+}\rangle + 4 \gamma \langle S_x \rangle \langle S_z \rangle, \\
\frac{d}{dt}\langle S_y\rangle &= -2 v\langle S_z \rangle -2\gamma\langle [S_z,S_y]_{+}\rangle + 4 \gamma \langle S_y \rangle \langle S_z \rangle, \\
\frac{d}{dt}\langle S_z\rangle &= 2v \langle S_y \rangle -4\gamma \langle S^{2}_z \rangle + 4 \gamma \langle S_z \rangle^2. 
\end{split}\; 
\label{IV:Ehrenfest1}\end{equation}

For coherent states, the expectation value of the anti-commutators in 
Eq.~(\ref{IV:Ehrenfest1}) can be simplified using the 
relation~\cite{GraefePRL2008}, \cite{GraefeJPA2010} 
\begin{equation}
\langle [S_i,S_{j}]_{+}\rangle = 
   \left(2-\frac{1}{S} \right)\langle S_i\rangle\; 
   \langle S_j\rangle + \delta_{ij}\, S\; .
\end{equation}
Thereby we obtain
\begin{equation}
\begin{split}
\dot{s}_x &= 2\gamma s_x s_z/S, \\
\dot{s}_y &= -2v s_z + 2\gamma s_y s_z/S,\\
\dot{s}_z &= 2v s_y + 2\gamma s^{2}_z/S- 2\gamma S,
\end{split}\; 
\label{SEE}
\end{equation}
where $s_j = \langle S_j  \rangle$. With the above equations it is easy to show that
\begin{equation}
\frac{d}{dt}\, \big (\, s_x^2 + s_y^2 + s_z^2\, \big ) = 0 \; ,
\label{SumExpVal}
\end{equation}
which means in particular that a coherent state remains coherent throughout
the evolution of the system (see App.~\ref{sec:App1}).

The analytical solution for this
system of equations is derived in App. \ref{AppSOL}. At the heart of this solution, there is the finding that the trajectories 
are circles on the sphere, which can be defined as the intersection of the 
sphere with a plane parallel to the $z$-axis and crossing the $y$-axis at a 
point $y_0/S = v/\gamma > 1$. In the $xy$ plane we can therefore draw a line 
through $(0,y_0)$ and the starting point of the trajectory and thereby find a 
second intersection point (of the line with the unit circle). Depending on the 
size of $\gamma$ this second point may be on one side or on the ``other side'' 
of the starting point, which explains the movement of the circle centers as 
$\gamma$ is increased.

\begin{figure}
  \centering
  \includegraphics[width= 1\textwidth]{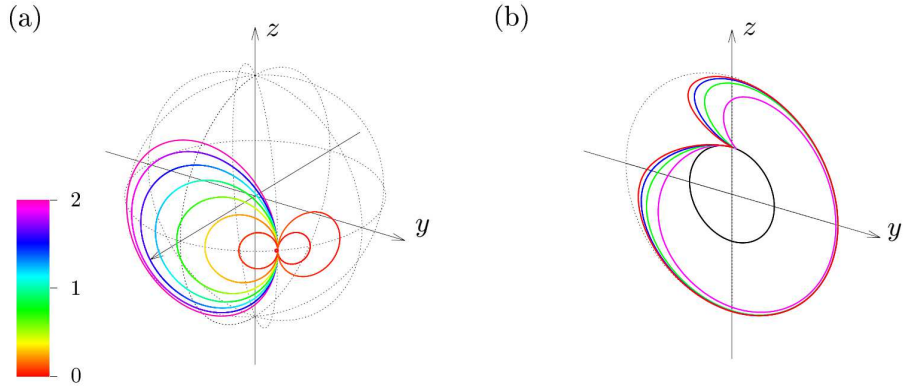}
  \caption{Trajectories of the expectation value of $\vec{S}$ for different 
values of $\gamma$. In panel (a) for the initial coherent state 
$\left|\theta = \nicefrac{\pi}{2} , \phi = \nicefrac{\pi}{4} \right\rangle$,
and for $\gamma = 0$, $0$.$1$, $\ldots$, $0$.$9$. from left (violet) to right 
(red). The line color represents the velocity on a common scale from 
0 (red) to 2 (violet). 
In panel (b) for the initial Dicke state $|S=10,m=4\rangle$ for $\gamma = 0$
(black solid line), $0.2$ (magenta), $0.4$ (green), $0.6$ (blue), and $0.8$ 
(red solid line).
The dotted line represents the circle $y^2 + z^2 = S^2$.}
\label{figtray}
\end{figure}

\paragraph{\bf Numerical results.} One way to study the dynamics of the evolution is through the expectation value of $\vec{S}$, or in components $\left( s_x, s_y, s_z \right)$.
Fig. \ref{figtray} shows the trajectories of Eq. (\ref{vecS}) 
for the initial coherent state 
$\left| \theta_0 = \nicefrac{\pi}{2}, \phi_0 = \nicefrac{\pi}{4} \right\rangle$ (a)
and for the initial Dicke state $|S=10,m=4\rangle$ (b) for different values of 
$\gamma$. According to Eq.~(\ref{SumExpVal}), for an initial coherent state, the time evolution of the  expectation value of $\vec{S}$ lies always on the surface of the sphere $\mathbb{S}^2$. In this case, it  represents the center of the distribution. Toghether with the change of ``sides'' of the circles, also we see that their radii decrease as $\gamma$ approaches $\sin \phi_0$. For completeness we have included on Fig. \ref{figtray}(a) the velocities of the different $\gamma$ trajectories. As stated earlier (Sec.~\ref{TDQD}), the distribution moves faster near the point $\left( \theta = \nicefrac{\pi}{2}, \phi = - \nicefrac{\pi}{2} \right)$. An explanation for
this speed up / slow down effect can be given by analyzing the disentangling
functions in Eq. (\ref{disec}). We recall that $h(t)$
is a real function and its effect is precisely the rotation of the
distribution. The slope of $h(t)$ defines the velocity for which the
distribution rotates. As Fig. \ref{fig:2} shows, we see that when $g(t)$ takes
values around its maxima, the local slope of $h(t)$ decreases; when $g(t)$
decreases, the local slope of $h(t)$ is more pronounced. 

On the other hand, the corresponding $\vec{s}$ values for the evolution of an initial Dicke state never reach the surface of $\mathbb{S} ^2$. We see that as $\gamma$ increases, the trajectories get closer to the circle $y^2 + z^2 = S^2$, though they never exactly reach it.

Additionally, the width of the distribution is given by the variances
\begin{equation}
  \Delta^2 S_j \left( t \right) = \left\langle S_j^2 \left( t \right) \right\rangle - \left\langle S_j \left( t \right) \right\rangle^2,
\end{equation}

\noindent where $j=x,y,z$. We compute the sum of the different variances for the evolution of the initial coherent state $\left| \theta_0 = \nicefrac{\pi}{2}, \phi_0 = \nicefrac{\pi}{4} \right\rangle$ and found 
\begin{equation}
  \sum_j \Delta^2 S_j(t) = S.
  \label{sumvartext}
\end{equation}

\noindent Notice that the time evolution of the sum  of variances is the same as in the standard coherent state 
Eq. (\ref{sumvar}), namely the time evolution leaves invariant (except for
the trace) the properties of the initial coherent state. 



\begin{figure}
  \centering
  \hspace*{-4mm}
  \includegraphics[scale=0.56]{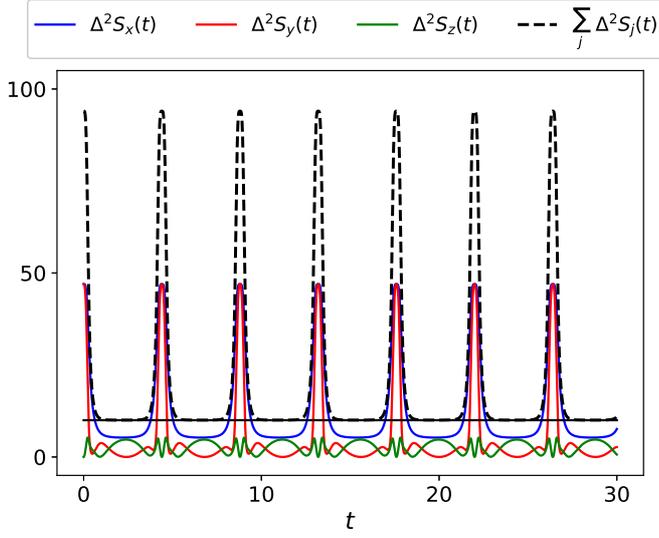}
  \caption{Time evolution of  the variances $\Delta^2 S_j(t)$  for the initial Dicke state $|S=10,m=4\rangle$ with $\gamma = 0.7$. We also plotted the sum $\sum_j \Delta^2 S_j(t)$ and the value $S=10$ (solid black line) for comparison.}
  \label{DPM}
\end{figure}

\noindent For initial Dicke states the variances behave differently. We plotted them in Fig.~\ref{DPM} for $|S=10,m=4\rangle$ and $\gamma = 0$.$7$. We see that the initial state has variance $\sim S^2$,
and periodically returns to that value. Interestingly, the sum $\sum_j \Delta^2 S_j(t)$ is almost equal to $S=10$ in periodic time intervals. Thus, the sum of fluctuations almost fulfills Eq.~(\ref{sumvar}) and the evolved Dicke state has periodically the properties of a coherent state: for certain periods of time the state attains a pseudo-coherent shape, as in Fig.~\ref{Dicke2} (though a numerical closer look shows that the sum of fluctuations never reaches exactly the value of $S$).

\noindent As an important final observation, we note that one can obtain a consistent evolution equation for the normalized Husimi function. Defining
\begin{equation}
  \rho' = \frac{\rho}{\operatorname{tr} \rho}
\end{equation}
one gets a modified version of Eq.~(\ref{eq:rho}), 
\begin{equation}
  i \partial_t \rho' = [H_0,\rho'(t)] + i[\Gamma, \rho'(t)]_{+} - i \frac{\rho'}{\operatorname{tr} \rho} \frac{d \operatorname{tr} \rho}{dt}.
\end{equation}
Using Eq.~(\ref{dtraza}),
\begin{equation}
  i \partial_t \rho' = [H_0,\rho'(t)] + i[\Gamma, \rho'(t)]_{+} + 4i\gamma \rho' \left\langle S_z \right\rangle.
\end{equation}
Thus for the Husimi function, since $ \partial_t{Q'}(\theta,\phi) = \partial_t{Q}_{\rho'}(\theta,\phi)$, one obtains
\begin{equation}
  \begin{split}
    \partial_t{Q}_{\rho'}(\theta,\phi) &=
 -2 (v \hat{l}_x +2 \gamma S\cos\theta -\gamma \sin\theta \partial_\theta\\&  -2i\gamma \rho' \left\langle S_z \right\rangle)Q'(\theta,\phi),
  \end{split}
\end{equation}
which leads to
\begin{eqnarray}\label{cleq1}
  \dot{\theta} &= &-2 v \sin\phi + 2\gamma \sin\theta, \\ \notag
  \dot{\phi} &= &-2v \cot\theta \cos\phi,\\ \notag
  \dot{Q'} &=& 4\gamma \left(S \cos\theta - \left\langle S_z \right\rangle \right).
\end{eqnarray}
As we pointed out, the evolution of a coherent state remains a coherent state. This is consistent with the latter equation, since in this case,
\begin{equation}
  \left\langle S_z \right\rangle = S \cos\theta,
\end{equation}
therefore
\begin{equation}
  \dot{Q'} = 0.
\end{equation}

\vspace{5mm}

\section{Conclusions}
\label{sec:Conclusions}

In this contribution we study the time evolution of a quantum system with
non-Hermitian, $\mathcal{PT}$-symmetric Hamiltonian with $\SOT$ as the dynamical
symmetry group, where the corresponding classical phase space is the Bloch 
sphere. We consider the simplest non-trivial Hamiltonian (linear in the 
angular momentum operators) and analyze the evolution of, both,
the Husimi representation and the
evolution of the expectation values of the angular momentum operators. 

The system has one single exceptional point (EP), where all eigenstates and
eigenvalues coalesce at the same time. Because of this, standard numerical or
analytical methods based on the diagonalization of the Hamiltonian quickly
become unstable. We overcome this problem by disentangling the
evolution operator into three different rotations, two of which are
``imaginary''.

In the Hermitian case, the dynamics in phase space would consist of
a rigid rotation around a fixed axis. Classical and quantum
dynamics would essentially agree, the evolution equation for the 
Husimi distribution function being equal to the Liouville equation for 
classical phase space densities. The evolution of expectation 
values of the angular momentum operators would also agree with the evolution
equation of the corresponding classical observables.

For the Hamiltonian studied here, many of these properties do no longer hold.
The evolution of the Husimi distribution may still be solved by the method of
characteristics, but the system of differential equations which 
determines the charactersitics is now quadratic and the characteristics 
themselves are deformed. In addition, the derivative of the Husimi function
along these characteristics is no longer constant. Hence, 
the norm of the evolving quantum state (i.e. its trace) is no longer
conserved. The phase space can be divided into the northern hemisphere
($\theta<\pi/2$) which acts as a source and the southern hemisphere ($\theta>\pi/2$) which acts as a sink.

Now, it is no longer possible to derive a closed evolution equation for the 
above mentioned expectation values. In addition since the Hamiltonian is linear
the Planck constant drops out from the Schr\" odinger equation, so that 
there is no semiclassical regime, and no valid Ehrenfest equations 
apparently making it impossible to identify classical trajectories. For instance,
studying the evolution of initial Dicke states, we find that the evolution of
the angular momentum expectation values is limitted to the $yz$-plane inside 
the Bloch sphere, while the corresponding evolution for initial coherent states
takes place on the Bloch sphere. This makes it possible to find different
initial quantum states, such that the trajectories trazed out by the 
expectation values cross.

Coherent states offer an elegant solution to this problem. As it turns out,
these states
remain coherent throughout their evolution. Hence the corresponding expectation
values trace a trajectory on the Bloch sphere. These trajectories agree with
the characteristics and thereby they provide a means to define a corresponding
classical dynamical system in a meaningful way.
Nevertheless, even a coherent state
will suffer from variations of its norm. Hence, the corresponding classical 
dynamics formulated in terms of the Liouville equation, really describes an
ensemble of particles whose size varies according to the particle density
function passing over the sources and sinks in phase space.

We have also studied the time
evolution under a similarity transformation that effectively diagonalizes the
Hamiltonian. In this case the similarity transformation of a coherent state
remains coherent, while the similarity transformation of a Dicke state
approaches a coherent state (the closer it is, the smaller $v-\gamma>0.9$).
After this, the
time evolution is a simple rotation around the $x-$axis.

For the present work, the disentangling procedure has been
essential and it will be interesting to investigate the possibility
to apply that method to PT-symmetric variantes of non-linear
models, such as the Kerr or the Lipkin-Meshkov-Glick models~\cite{valtierra2020quasiprobability}.


\vspace*{5mm}

\appendix

\section{Properties of coherent and Dicke states}
\label{sec:App1}

\noindent We are using as initial states two states with different properties of their distributions: one that is localized and one that is not. In fact, it is well known \cite{Klimovbook2009} that for a coherent state $\left| \theta_0 , \phi_0\right\rangle$, centered at $\left( \left\langle S_x \right\rangle , \left\langle S_y \right\rangle , \left\langle S_z \right\rangle \right)$ with
\begin{eqnarray}
\left\langle S_x \right\rangle & = & S \sin \theta_0 \cos \phi_0, \nonumber\\ 
\left\langle S_y \right\rangle & = & S \sin \theta_0 \sin \phi_0, \nonumber\\
\left\langle S_z \right\rangle & = & S \cos \theta_0,
\label{promch}
\end{eqnarray}

\noindent its variances are,
\begin{eqnarray}
\Delta^2 S_x & = & \frac{S}{2} \left( 1- \sin^2 \theta_0 \cos^2 \phi_0 \right), \nonumber\\
\Delta^2 S_y & = & \frac{S}{2} \left( 1- \sin^2 \theta_0 \sin^2 \phi_0 \right), \nonumber\\
\Delta^2 S_z & = & \frac{S}{2} \left( 1- \cos^2 \theta_0 \right),
\label{varch}
\end{eqnarray}

\noindent with the property
\begin{equation}
\Delta^2 S_x + \Delta^2 S_y + \Delta^2 S_z = S.
\label{sumvar}
\end{equation}

\noindent So the state is localized because its varaiances are $\sim S$.

\noindent The Dicke state $\left| S, m\right\rangle$, fulfills
\begin{equation}
\left\langle S_x \right\rangle = 0, \left\langle S_y \right\rangle = 0, \left\langle S_z \right\rangle = m,
\label{promd}
\end{equation}

\noindent with variances
\begin{eqnarray}
\Delta^2 S_x & = & \frac{1}{2} \left( S^2 + S - m^2 \right), \nonumber\\
\Delta^2 S_y & = & \frac{1}{2} \left( S^2 + S - m^2 \right), \nonumber\\
\Delta^2 S_z & = & 0,
\label{vard}
\end{eqnarray}

\noindent one can see that the state is not localized, i.e. its variances are $\sim S^2$, when $m^2 \sim S$.

Furthermore, consider the functional,
\[ \mathcal{L} \quad :\quad |\Psi\rangle \to \langle S_x\rangle^2 
  + \langle S_x\rangle^2 + \langle S_x\rangle^2\; , \]
then we want to show two things: (i) In any case $\mathcal{L}[\Psi] \le S$;
(ii) $\mathcal{L}[\Psi] = S$ if and only if $\Psi$ is a coherent state. Both
statements follow rather immediately from the invariance of $\mathcal{L}$ under
rotations. For instance to prove (i) assume there exits a $\Psi$ with 
$\mathcal{L}[\Psi] > S$ then we can find a rotation which makes 
$\langle S_x\rangle = \langle S_y\rangle = 0$. Hence, we find 
$\langle S_z\rangle > S$ which is impossible because of the eigenvalues of 
$S_z$.

To prove (ii) we note that one direction of the ``if and only if'' is clear:
If $\Psi$ is a coherent state then $\mathcal{L}[\Psi] = S$.
To show the other case, assume $\mathcal{L}[\Psi] < S$ and do the rotation
just as in the previous case. Then we find $\langle S_z\rangle < S$, which
means that the state in question cannot be the eigenstate $|S,S\rangle$ which
it should be (if $\Psi$ really were a coherent state, according to Eq.~(\ref{edocoh})).

\section{Evolution Equation on Phase Space} \label{EvoPS}

\noindent The von Neumann equation can be written as
\begin{equation}
i\partial_{t}\rho (t) = [H_0,\rho(t)]+ i [\Gamma,\rho(t)]_{+}.
\end{equation}

\noindent To recast this equation on phase space, we multiply by the kernel 
and take the trace
\begin{eqnarray}\label{eq3}
i\partial_{t}\mathrm{Tr}\Big(\rho (t) \hat{\omega}_{Q}(\theta,\phi)\Big) & = &  \mathrm{Tr}\Big([H_0,\rho(t)]\hat{\omega}_{Q}(\theta,\phi)\Big)+ \nonumber \\
& & i \mathrm{Tr}\Big([\Gamma,\rho(t)]_{+}\hat{\omega}_{Q}(\theta,\phi)\Big).
\end{eqnarray}
The correspondence rules or Bopp operators are very useful \cite{ZFG,ACGT,Klimovbook2009}: 
\begin{equation}\label{cr1}
\left.
\hat{\rho}\hat{S}_z \atop
\hat{S}_z\hat{\rho}
\right\}\longleftrightarrow    \left\{ \Big(\mp \frac{1}{2}l_z + \Lambda_0(\theta,\phi)  \right. \Big)\hat{\rho},
\end{equation}

\begin{equation}\label{cr2}
\left.
\hat{\rho}\hat{S}_\pm \atop
\hat{S}_\pm\hat{\rho}
\right\}\longleftrightarrow    \left\{ \Big(\mp \frac{1}{2}l_\pm + \Lambda_\pm(\theta,\phi)  \right. \Big)\hat{\rho},
\end{equation}

\noindent where $l_{\pm,z}$ are the first order differential operators,
\begin{equation}
l_{\pm}= e^{\pm i \phi}\left(\pm \partial_{\theta}+ i\mathrm{cot}\theta \partial_{\phi}   \right), \quad l_z=-i\partial_{\phi}.
\end{equation}

\noindent The operators $\Lambda_{0,\pm}(\theta,\phi)$ are
\begin{equation}
\Lambda_{0}(\theta,\phi) = \frac{1}{2} \left(\frac{1}{\epsilon}\cos\theta -\cos\theta -\sin\theta \partial_{\theta}   \right),
\end{equation}

\begin{equation}
\Lambda_{\pm}(\theta,\phi)= \frac{e^{\pm i\phi}}{2\epsilon}\pm \frac{1}{2}\Big[\cos\theta l_{\pm}-e^{\pm i\phi}\sin\theta (l_z \pm 1)    \Big],
\end{equation}

\noindent with $\epsilon =(2S+1)^{-1}$.

\noindent Using Eqs.~\eqref{cr1} and \eqref{cr2}, Eq.~\eqref{eq3} takes the following form
\begin{equation}
i\partial_{t}Q(\theta,\phi) = -2\left( v l_x +2i\gamma S \cos\theta - i\gamma \sin\theta \partial_{\theta}\right)Q(\theta,\phi).
\label{evoQ}
\end{equation}

Another way to obtain Eq.~(\ref{evoQ}) is through \cite{KlimovJMP2002}. Eq.~(\ref{eq3}), can be recasted in the following form
\begin{equation}
i\partial_t Q(\theta,\phi) = \frac{i}{S} \{Q(\theta,\phi),Q_{H_0}(\theta,\phi)\}_p +  i \ \hat{\Xi} (\theta,\phi)Q(\theta,\phi).
\label{eq:Q}
\end{equation}

\noindent Here 
\begin{equation}
\{F,G\}_{p} = \frac{1}{\sin\theta}\left(\partial_{\phi}F \ \partial_{\theta}G - \partial_{\theta}F \ \partial_{\phi}G   \right)
\end{equation}

\noindent are the Poisson brackets on the sphere and 
\begin{equation}
\hat{\Xi} (\theta,\phi)Q(\theta,\phi) = \operatorname{Tr}([\Gamma, \rho(t)]_+ \hat{\omega}_Q \left( \theta, \phi\right))
\end{equation}

\noindent is the corresponding action of the anticommutator 
of $\Gamma$ and the density matrix on $Q$. 

\noindent The structure of Eq.~(\ref{eq:Q}) comprises a sympletic structure given by
the Possion brackets $\lbrace , \rbrace_p$ plus a term which originates from
the $\mathcal{PT}$-symmetric structure of the Hamiltonian. This structure is 
similar to the one found by Graefe and coworkers in \cite{GraefeJPA2010}.

\section{Analytical solution of the Ehrenfest Equations for coherent states}
\label{AppSOL}

\noindent According to Eq.~(\ref{SEE}), let $x$, $y$ and $z$ be $s_x/S$, $s_y/S$ and $s_z/S$, respectively. Then we get the system of equations,
\begin{eqnarray}
\dot{x} & = & 2\gamma x z \notag \\
\dot{y} & = & -2v z + 2\gamma y z\notag \\
\dot{z} & = & 2v y + 2\gamma z^{2}- 2\gamma,
\label{sdeq}
\end{eqnarray}

\noindent valid for the evolution of coherent states. Our aim is it to find the trajectories on the sphere. First, we divide the second equation by the first and obtain
\begin{equation}
\frac{dy}{dx} = y'(x) = \left(y-\frac{v}{\gamma} \right) \frac{1}{x} \ \Longrightarrow \ y(x) = C_1 x + \frac{v}{\gamma}.
\end{equation}

\noindent Here $C_1$ is a first integration constant, which is determined from the initial conditions and yields
\begin{equation}
C_1 = -\frac{v/\gamma - y_0}{x0} = -\frac{v/\gamma - \sin \theta_0 \sin \phi_0}{\sin \theta_0 \cos \phi_0},
\end{equation}

\noindent where $\theta_0$ and $\phi_0$ are the parameters of the initial coherent state. Now we simply use the fact that (Eq. (\ref{SumExpVal}))
\begin{equation}
x^2 + y^2 + z^2 = 1.
\end{equation}

\noindent This means tha the trajectory is just the intersection of the unit sphere with the plane. This always gives a circle, with its center in the $(x, y)$ plane. Let $(x_c, y_c)$ be the center of that circle and $(\Delta x, \Delta y)$ be such that
\begin{equation}
\vec{r}(\alpha) = \left(\begin{array}{c}
x_c \\ y_c \\ 0
\end{array} \right) + \left(\begin{array}{c}
\Delta x \\ \Delta y \\ 0
\end{array} \right) \cos \alpha + \left(\begin{array}{c}
0 \\ 0 \\ \sqrt{\Delta x^2 + \Delta y^2}
\end{array} \right) \sin \alpha.
\end{equation}

\noindent Then we can calculate all the unknowns $x_c$, $y_c$, $\Delta x$ and $\Delta y$ from the two intersection points, $x_{1,2}$, of the line $y = C_1 x +v/\gamma$ and the circle $x^2 + y^2 = 1$ in the $(x, y)$ plane:
\begin{equation}
x^2 + \left( C_1 x +v/\gamma \right)^2 = 1,
\end{equation}

\noindent then
\begin{eqnarray}
x_{1,2} & = & -p \pm \sqrt{p^2 - q}, \nonumber \\
p & = & \frac{v}{\gamma} \frac{C_1}{1+C_1^2}, \nonumber \\
q & = & \frac{v^2/\gamma^2 - 1}{1 + C_1^2}.
\end{eqnarray}

\noindent With this we obtain
\begin{eqnarray}
x_c & = & \frac{x_1 + x_2}{2} = -p, \nonumber \\
y_c & = & C_1 x_c +v/\gamma = -C_1 p +v/\gamma,
\end{eqnarray}

\noindent and also
\begin{eqnarray}
\Delta x & = & \frac{x_1 - x_2}{2} = \sqrt{p^2 - q}, \nonumber \\
\Delta y & = & C_1 \Delta x = C_1 \sqrt{p^2 - q}.
\label{deltaxy}
\end{eqnarray}

\noindent Now, as we observed some trajectories that degenerated to a point, let us find a solution for
\begin{eqnarray}
\Delta x & = & \Delta y = 0 \nonumber \\
\Delta z & = & \sqrt{\Delta x^2 + \Delta y^2} = 0.
\end{eqnarray}

\noindent It is clear from Eq.~(\ref{deltaxy}) that this is fulfilled when $p^2 = q$, which yields
\begin{equation}
\gamma = \frac{\pm i \cos \theta_0 \cos \phi_0 + \sin \phi_0}{\sin \theta_0} v .
\end{equation}

\noindent We see that real solutions exist only for $\theta_0 = \pi/2$. Only the trajectory of the evolution of an initial coherent state located on the $(x, y)$ plane degenerates to a point. The value of $\gamma$ that leads to this is ($v = 1$)
\begin{equation}
\gamma = \sin \phi_0.
\label{gammapunto}
\end{equation}

\noindent Furthermore, in this case ($\theta_0 = \pi/2$, $v=1$) we have
\begin{eqnarray}
x_c - x_0 & \sim & \sin \phi - \gamma , \nonumber \\
y_c - y_0 & \sim & \sin \phi - \gamma ,
\label{cambsign}
\end{eqnarray}

\noindent so both, $x_c - x_0$ and $y_c - y_0$, suffer a change of sign before and after Eq.~(\ref{gammapunto}). This means that trajectories for $\gamma < \sin \phi_0$ and for $\gamma > \sin \phi_0$ are centered in different sides of the common point ($\theta_0, \phi_0$). All this is exemplified in Fig.  \ref{figtray}.

\subsection*{Acknowledgements}
The authors are grateful to A.~B. Klimov for many fruitful discussions. T.G. received financial support from CONACyT through the grant ``Ciencia de Frontera
2019'', No. 10872.

%
\section{Declarations}
\subsection*{Funding}

CONACyT ``Ciencia de Frontera 2019'', Number 10872.

\subsection*{Conflict of interest}

The authors declare that they have no conflict of interest.

\subsection*{Availability of data and material}

The data generated for the paper are available upon personal
request to the authors.

\subsection*{Code availability}

The numerical programs are available upon personal request to the authors.

\subsection*{Authors' contributions}

All authors have contributed equally to the research.

%


%
%

\end{document}